# Evolution of galaxies in clusters

**Bianca Poggianti**[*][†]

*INAF-Padova Astron. Observatory, Italy*
*E-mail:* `poggianti@pd.astro.it`

I summarize what is know about the evolution of galaxies in clusters from the observational point of view presenting results at high ($z \sim 1$), intermediate ($\sim 0.5$) and low ($z \sim 0$) redshifts. I comment on the comparison between observations and predictions of CDM models, highlighting the observational landmarks more relevant for this comparison, such as the establishment and evolution of the morphology-density relation, the Butcher-Oemler effect, the evolution of red galaxies/ellipticals, the star formation histories of galaxies in clusters, the downsizing effect and the history of mass assembly.

*Baryons in Dark Matter Halos*
*5-9 October 2004*
*Novigrad, Croatia*

[*]Speaker.
[†]The author wishes to thank the organizers of this School+meeting for their kind invitation and generous support, and for unlimited patience in waiting for this written contribution.

## 1. Introduction

The goal of this lecture is to give an overview of the state-of-the-art in the field of galaxy evolution in clusters and to point out the main observational milestones that a model of galaxy formation and evolution should attempt to reproduce and explain, trying to underline where the current models succeed and where they most struggle.

Understanding the role of *environmental conditions* in determing how galaxies are and evolve is *especially important in the context of a hierarchical cosmological scenario.* As structures grow, galaxies can join more and more massive structures and experience different environmental conditions during their history. Separating galaxy evolution into "field" and "cluster" evolution hardly makes sense in this scenario, since a cluster galaxy today may have been a group or field galaxy during the earlier phases of its evolution.

The physical mechanisms that are usually considered when trying to assess the influence of the environment on galaxy evolution can be grouped in four main families:

1. Mergers and strong galaxy-galaxy interactions (Toomre & Toomre 1972, Hernquist & Barnes 1991, see Mihos 2004 for a review). They are most efficient when the relative velocities between the galaxies are low, thus are expected to be especially efficient in galaxy groups.

2. Tidal forces due to the cumulative effect of many weaker encounters (also known as "harassment") (Richstone 1976, Moore et al. 1998). They are expected to be especially important in clusters, and particularly on smaller / lower mass galaxies.

3. Gas stripping - Interactions between the galaxy and the inter-galactic medium (IGM) (Gunn & Gott 1972, Quilis et al. 2000). The interstellar medium of a galaxy can be stripped via various mechanisms, including viscous stripping, thermal evaporation and – the most famous member of this family – ram pressure stripping. Ram pressure can be efficient when the IGM gas density is high and the relative velocity between the galaxy and the IGM is high. These conditions are expected to be met especially in the very central regions of cluster cores.

4. Strangulation (also known as starvation, or suffocation) (Larson, Tinsley & Caldwell 1980, Bower & Balogh 2004). Assuming galaxies possess an envelope of hot gas that can cool and feed the disk with fuel for star formation, the removal of such reservoir of gas is destined to inhibit further activity once the disk gas is exhausted. In semi-analytic models, for example, the gas halo is assumed to be removed when a galaxy enters as satellite in a more massive dark matter halo.

Note that while stripping gas from the disk induces a truncation of the star formation activity on a short timescale ($\sim 10^7$ yrs), strangulation is expected to affect a galaxy star formation history on a long timescale ($> 1$ Gyr) provoking a slowly declining activity which consumes the disk gas after the supply of cooling gas has been removed.

Some of these processes have en empirical motivation (e.g. ram pressure stripping and mergers can be observed at work), while others have a more "theoretical ground" (e.g. strangulation and harassment).

The former two of these families of processes affect the galaxy structure, thus morphology, in a direct way: the merger of two spirals can produce an elliptical galaxy, and repeated tidal encounters can change a late–type into an early-type galaxy. The latter two families, instead, act on the gas content of galaxies, hence their star formation activity, and can modify their morphologies in an indirect way: once star formation is halted in a disk, this can fade significantly, the bulge-to-disk relative importance can change and the galaxy appearance and morphology can appear significantly modified.

## 2. Evolution with redshift

### 2.1 Galaxy morphologies and star formation activity

At the time of this School, high–quality data of distant clusters have been obtained with the Advanced Camera for Surveys by several groups, and results should appear soon (Postman et al. 2005, Desai et al. in prep.). Here I summarize the published results, which are all based on Wide Field and Planetary Camera 2 data.

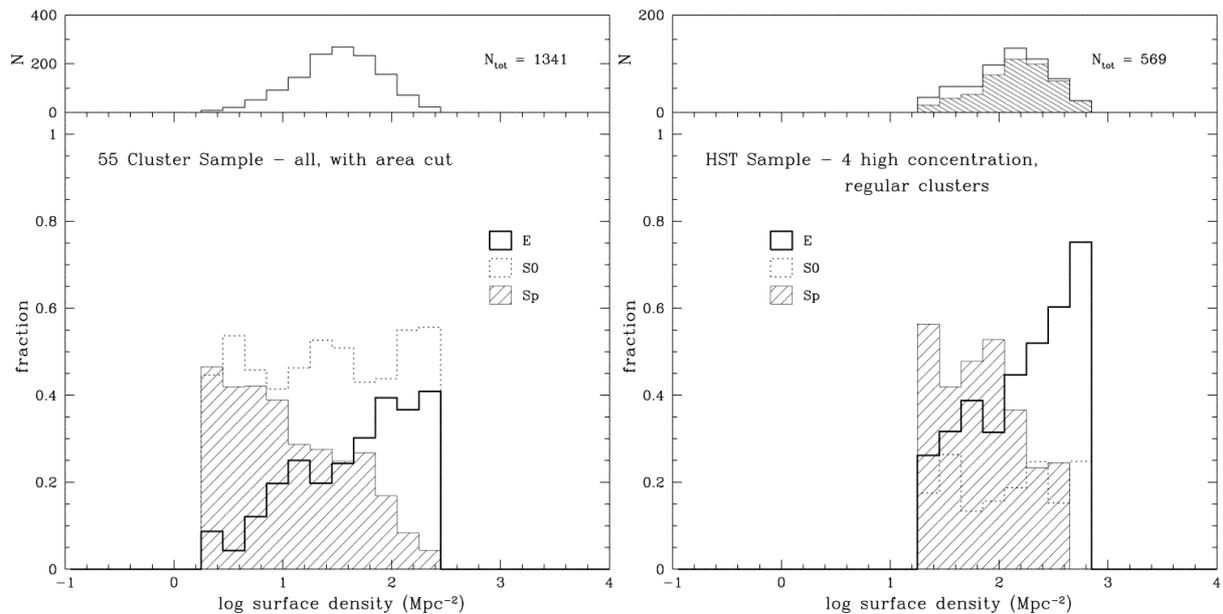

**Figure 1:** From Dressler et al. (1997). *Left.* Morphology-density relation for 55 clusters at low redshift (Dressler 1980). *Right.* Morphology-density relation for 4 regular clusters at $z \sim 0.45$ (Dressler et al. 1997).

The HST images have revealed the presence of large numbers of spiral galaxies in all distant clusters observed, in proportions that are much higher than in nearby clusters of similar richness (Dressler et al. 1994, Couch et al. 1994, Dressler et al. 1997). This is reflected also in the evolution of the Morphology-Density (MD) relation. The MD relation is the observed correlation between the frequency of the various Hubble types and the local galaxy density, normally defined as the projected number density of galaxies within an area including its closest neighbours. In clusters in the local Universe, the existence of this relation has been known for a long time: ellipticals are frequent in high density regions, while the fraction of spirals is high in low density regions (Oemler 1974, Dressler et al. 1980). At $z = 0.4 - 0.5$, an MD relation is already present, but it is *quantitatively* different from the relation at $z = 0$: the fraction of S0 galaxies at $z = 0.5$ is much lower, at all densities, than in clusters at $z = 0$ (Fig. 1, Dressler et al. 1997). The fraction of S0s in clusters appears to increase towards lower redshifts, while the proportion of spirals correspondingly decreases (Fig. 2, Dressler et al. 1997, Fasano et al. 2000). Interestingly, ellipticals are already as abundant at $z = 0.5$ as at $z = 0$. These findings strongly suggest that a significat fraction of the S0 galaxies in clusters today have evolved from spirals at relatively recent epochs. Adopting a more conservative distinction between "early-type" (Es+S0s) and late-type (spirals) galaxies, a similar evolution is found, with the early-type fraction decreasing at higher redshifts (van Dokkum et al. 2000, Lubin et al. 2002). First results at $z \sim 0.7 - 1.3$ seem to indicate that between $z = 0.5$ and $z = 1$ what changes in the MD relation is only the occurrence of early-type galaxies in the very highest density regions (Smith et al. 2004).

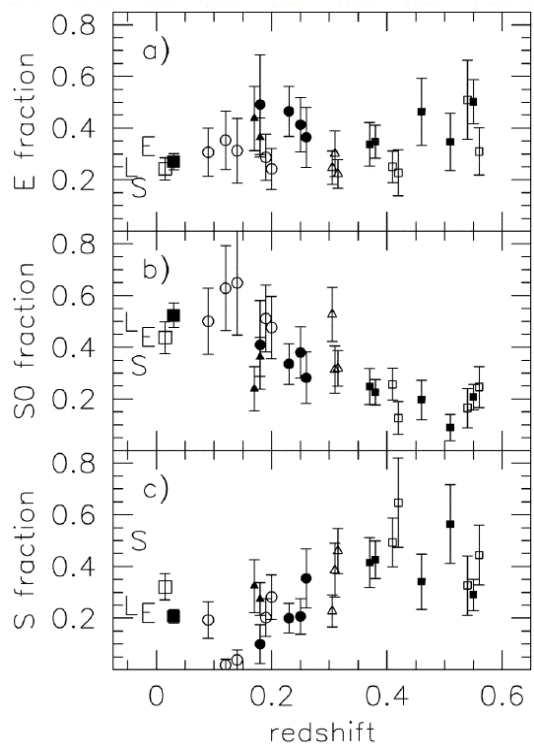

**Figure 2:** From Fasano et al. (2000). Evolution of the morphological mix in clusters. The fraction of ellipticals, S0 and spiral galaxies is shown for clusters between $z = 0.55$ and $z = 0$.

Interestingly, "an" MD relation, originally found in the cluster cores, has been shown to be present across a wide range of environments, in all types of clusters, rich and poor, concentrated and irregular, with low and high $L_X$ (Dressler 1980, Balogh et al. 2002a), in groups (Postman & Geller 1984) and at large cluster radii (Treu et al. 2003), though the relation seems to vary from rich to less rich clusters (Dressler 1980, Balogh et al. 2002).

Comparing HST morphologies and spectroscopy, it has been shown that galaxies with active star formation in distant clusters are for the great majority spirals (Dressler et al. 1999), but that the viceversa is not always true: several of the cluster spirals, in fact, do not display any emission line in their spectra, and both their spectra and their colors indicate a lack of current star formation activity (Poggianti et al. 1999, Couch et al. 2001, Goto et al. 2003). These "passive spirals" might be an intermediate stage when star–forming spirals are being transformed into passive S0 galaxies.

Historically, the first evidence for a higher incidence of star-forming galaxies in distant clusters compared to nearby clusters came from photometric studies that identified large numbers of *blue* galaxies (Butcher & Oemler 1978, 1984, Ellingson et al. 2001, Kodama & Bower 2001). Those galaxies in distant clusters that are *red* and already lie on the color-magnitude sequence have instead very old stellar populations formed at $z > 2-3$ that have evolved passively after that (Ellis et al. 1997, Kodama et al. 1998, Barger et al. 1998, van Dokkum et al. 1999,2000,2001). They are only a subset of the galaxies that lie on the red sequence today (van Dokkum & Franx 2001) but they are easily recognizable as signposts of high density regions out to very high redshifts (e.g. Stanford et al. 2002, Blakeslee et al. 2003, De Lucia et al. 2004). As we will see in Sec. 2.2, the fraction of galaxies involved in the evolution from blue to red doesn't depend only on redshift, but strongly also on galaxy mass.

Compared to colors, spectroscopy is a more direct way to identify galaxies with ongoing star formation. For distant galaxies, the H$\alpha$ line is redshifted at optical wavelengths that are severely affected by sky or is observed in the near-IR, thus the feature most commonly used is the [OII]$\lambda$3727 line.

In the MORPHS sample of 10 clusters at $z \sim 0.4 - 0.5$, the fraction of emission–line galaxies is $\sim 30\%$ for galaxies brighter than $M_V = -19 + 5\log h^{-1}$ (Dressler et al. 1999, Poggianti et al. 1999). In the CNOC cluster sample, at an average redshift $z \sim 0.3$, this fraction is about 25% (Balogh et al. 1999). This incidence is much higher than it is observed in similarly rich clusters at $z = 0$ (Dressler, Thompson & Shectman 1988). Significant numbers of emission-line galaxies have been reported in virtually all spectroscopic surveys of distant clusters (e.g. Couch & Sharples 1987, Fisher et al. 1998, Postman et al. 1998, 2001).

The increasing importance of the [OII] emission with redshift can also be assessed from cluster composite spectra, that are obtained summing up the light from all galaxies in a given cluster to produce a sort of "cluster integrated spectrum" (Fig. 3, Dressler et al. 2004). As expected, the strength of [OII] in these composite spectra displays a large cluster–to–cluster variation at any redshift, but there is a tendency for the $z = 0.5$ clusters to have on average a stronger composite EW([OII]) than the clusters at $z = 0$.

If numerous observations indicate that emission–line galaxies were more prominent in clusters in the past than today, and if these results are unsurprising given the evolution with $z$ of the star formation activity in the general "field", *quantifying* this evolution in clusters has proved to be very hard. The fact that the emission–line incidence varies strongly from a cluster to another at all redshifts, and the relatively small samples of clusters studied in detail at different redshifts, have so far hindered our progress in measuring how the fraction of emission–line galaxies evolves with redshift as a function of the cluster properties.

Discriminating between cosmic evolution and cluster–to–cluster variance is a problem also for H$\alpha$ cluster–wide studies. Using narrow–band imaging or multiplex multislit capabilities, a handful of clusters have been studied to date at $z \geq 0.2$ (Couch et al. 2001, Balogh et al. 2002b, Finn et al. 2004, and submitted, Kodama et al. 2004, Umeda et al. 2004). These studies have confirmed that the fraction of emission–line (H$\alpha$–detected, in this case) galaxies is lower in clusters than in the field at similar redshifts, and have shown that the bright end of the H$\alpha$ luminosity function does not seem to depend strongly on environment. As shown in Fig. 4, the number of clusters studied is still insufficient to pin down the star formation rate (SFR) per unit of cluster mass as a function of redshift AND of global cluster properties such as the cluster velocity dispersion.

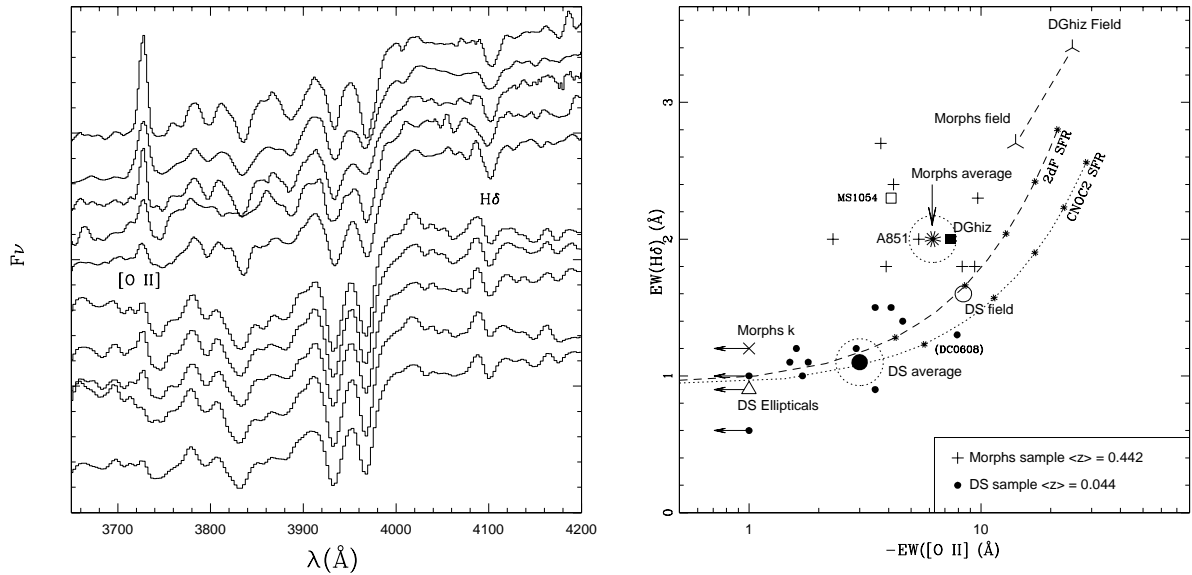

**Figure 3:** From Dressler et al. (2004). *Left.* Composite spectra of five clusters at $z \sim 0.5$ (top five) and five clusters at $z \sim 0$ (bottom five). The [OII] line is generally more prominent in the high-z spectra. *Right.* Equivalent widths of [OII] versus H$\delta$ as measured from composite spectra of clusters at $z \sim 0.4-0.5$ (crosses) and clusters at $z \sim 0$ (filled dots).

A word of caution is compulsory when using emission–lines and assuming they provide an unbiased view of the evolution of the star formation activity in cluster galaxies. There are several indications that dust extinction is in fact important and strongly distorts our view of the star formation activity in at least some cluster galaxies. Evidence for dust arises from optical spectroscopy itself, which finds many dusty starbursting or star–forming galaxies with relatively weak emission–lines both in distant clusters and in the field at similar redshifts (Poggianti et al. 1999, Shioya et al. 2000, Poggianti et al. 2001a, Bekki et al. 2001). The radio–continuum detection of galaxies with no optical emission lines (Smail et al. 1999, Miller & Owen 2002) and mid-IR estimates of the star formation rate (Duc et al. 2002, Biviano et al. 2004, Coia et al. 2005a,b) indicate that even the majority or all of the star formation activity of some cluster galaxies can be obscured at optical wavelengths. Whether taking into account dust obscuration changes significantly the evolutionary picture inferred from emission lines is still a critical open question that Spitzer is likely to answer.

Precious informations about the star formation histories of cluster galaxies can also be obtained from absorption–line spectra. In distant clusters, the presence of galaxies with strong Balmer lines in absorption in their spectra, and no emission lines, testifies that these are post-starburst/post-starforming galaxies observed soon after their star formation activity was interrupted and observed within 1-1.5 Gyr from the halting (Dressler & Gunn 1983, Couch & Sharples 1987, see Poggianti 2004 for a review). These galaxies have been found to be proportionally more numerous in distant clusters than in the field at comparable redshifts (Dressler et al. 1999, Poggianti et al. 1999, Tran et al. 2003, 2004). Their spectral characteristics and their different frequency as a function of the environment are a strong indication for a truncation of the star formation activity related to the dense environment.

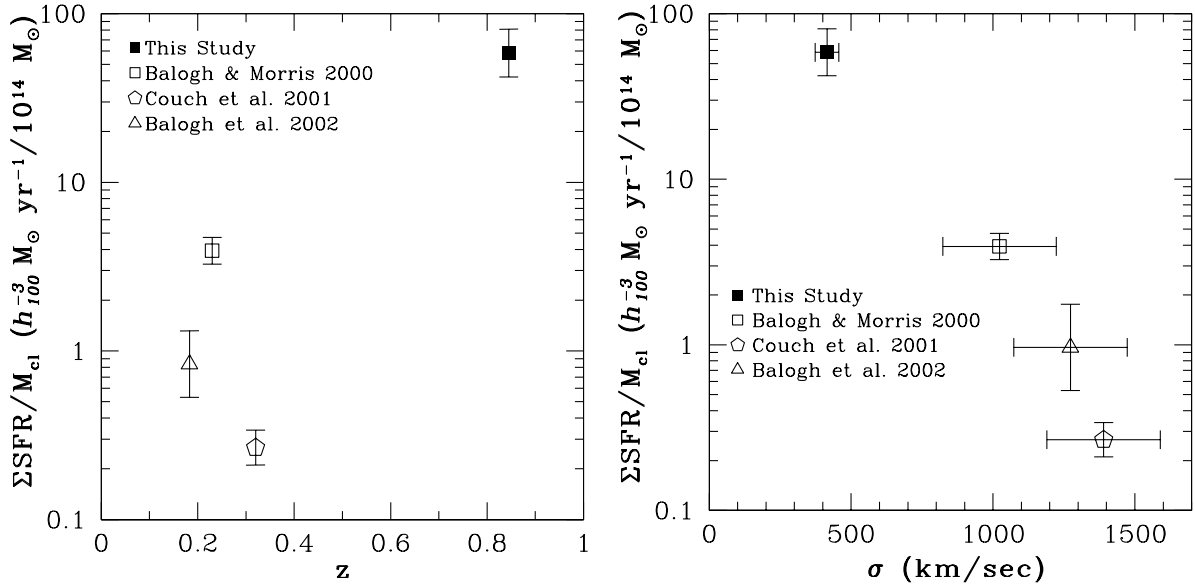

**Figure 4:** From Finn et al. (2004). Star formation rate per unit of cluster mass, as measured from Hα narrow-band imaging, as a function of redshift (left) and cluster velocity dispersion (right).

Alltogether, the findings described in this section suggest that many galaxies have stopped forming stars in clusters quite recently, as a consequence of environmental conditions switching off their star formation activity, and that (largely, in parallel) many galaxies have morphologically evolved from late to early type galaxies.

### 2.2 Dependence of the star formation history on galaxy mass - Mass assembly

How does the evolution depend on the galaxy mass? Does the environmental dependence of galaxy properties change with galaxy mass?

It has always been known that fainter, lower mass galaxies on average are bluer than higher mass galaxies, and on average have more active star formation. The references to old and new papers showing this could easily fill this review.

Recently, the interest in this well established observational result has grown, and its implications have been more and more appreciated. In *all* environments, lower mass galaxies have on average a more protracted star formation history. This implies that, on average, going to lower redshifts, the maximum luminosity/mass of galaxies with significant star formation activity progressively decreases. This "downsizing effect", observed in and outside of clusters, indicates an "anti-hierachical" history for the star formation in galaxies, which parallels a similar effect observed for AGNs (Cristiani et al. 2004, Shankar et al. 2004).

In clusters, innumerable results have shown the existence of a downsizing effect (Smail et al. 1998, Gavazzi et al. 2002, De Propris et al. 2003, Tran et al. 2003, De Lucia et al. 2004, Kodama et al. 2004, Poggianti et al. 2004, to name a few). A direct observation of this effect at high redshift is shown in Fig. 5 and illustrates the consequence of downsizing on the characteristics of the color-magnitude red sequence in clusters. A deficiency of faint red galaxies is observed compared to Coma in all four clusters studied, despite of the variety of cluster properties. The red luminous galaxies are already in place on the red sequence at $z \sim 0.8$, while a significant fraction of the faint galaxies must have stopped forming stars and, consequently, moved on to the red sequence at lower redshifts. A downsizing effect is also observed analyzing the post-starburst populations in clusters: the maximum mass of post-starburst galaxies evolves with z, being higher in distant clusters (Tran et al. 2003, Poggianti et al. 2004). The fraction of S0 galaxies with recent star formation in the Coma cluster is higher at fainter luminosities (Poggianti et al. 2001b), which is consistent with them being the descendants of typical star-forming spirals at intermediate redshift (see Sec. 2.1).

These results and many others have shown that the star formation history of a galaxy (in clusters and in the field) depends on average strongly on the galaxy mass (see also Kauffmann et al. 2003, 2004, Balogh et al. 2004b). Another important aspect is whether the *mass distribution* of galaxies varies with redshift. The stellar mass function, traced by the K-band luminosity function, has been shown to evolve very little *at the bright end* in clusters between $z = 1$ and $z = 0$: Kodama & Bower (2003) find an evolution which is consistent with passive evolution, and inconsistent with semi-analytic models that predict instead an increase in the characteristic mass of a factor $> 3$ over the same redshift range (Fig. 6). This result is in agreement with the evolution of $K^\star$ found in a cluster at $z = 1.2$ by Toft et al. (2004), who also find a strong evolution *of the faint end* of the K-band luminosity function: the fraction of low mass galaxies appears much lower in this cluster at $z = 1.2$ than in clusters at $z = 0$.

## 3. Local universe

In the previous sections we have dealt with the evolution of galaxy morphologies, star formation activity and masses. In this section I want to mention a couple of results established observationally at low redshifts, regarding galaxy properties and trends whose evolution with redshift is still unknown: the StarFormation-Density (SFD) relation, and the gas and star formation distributions within galaxies.

It has been known for a long time that in the nearby Universe also the average star formation activity correlates with the local density: in higher density regions, the mean star formation rate per galaxy is lower. This is not surprising, given the existence of the MD relation: the highest density regions have proportionally more early-type galaxies devoid of current star formation.

Interestingly, the correlation between mean SF and local density extends to very low local densities, comparable to those found at the virial radius of clusters, and such a correlation exists also outside of clusters (Lewis et al. 2002, Gomez et al. 2003; see also Gray et al. 2004 showing for the first time an interesting relation between the star formatin activity and the dark matter density field measured by weak gravitational lensing). Again, this seems to parallel the fact that an MD relation is probably existing in all environments, – though it is *not the same* in all environments, as discussed in Sec. 2.1.

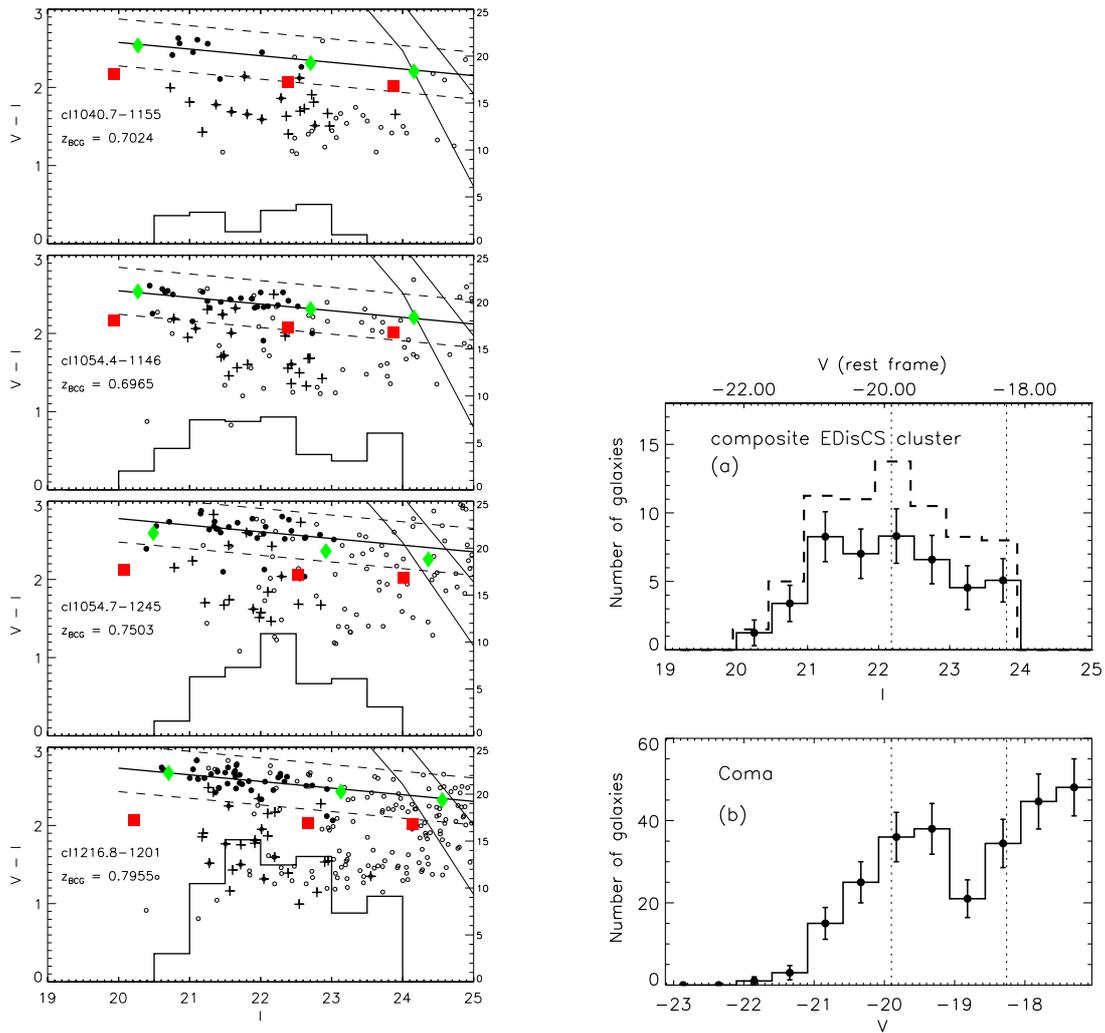

**Figure 5:** From De Lucia et al. 2004. *Left.* Color-magnitude diagrams of four clusters at $z = 0.7 - 0.8$ from the ESO Distant Cluster Survey. Histograms represent the magnitude distribution of galaxies within $3\sigma$ from the red sequence. *Right.* The magnitude distribution of red sequence galaxies at $z \sim 0.75$ (top) is compared with the one of red galaxies in the Coma cluster (bottom). High-z clusters exhibit a clear deficit of low-luminosity passive red galaxies compared to Coma and other nearby clusters.

A variation in the mean SF/galaxy with density can be due either to a difference in the *fraction of star-forming galaxies*, or in the *star formation rates* of the star-forming galaxies, or a combination of both. In a recent paper, Balogh et al. (2004a) have shown that the distribution of H$\alpha$ equivalent widths (EW) in star-forming galaxies does not depend strongly on the local density, while the fraction of star-forming galaxies is a steep function of the local density, in all environments. Again, a dependence on the *global environment* is observed, in the sense that, at a given local density, the fraction of emission-line galaxies is slightly lower in environments with high density on large scales ($\sim 5$ Mpc) (but see Kauffmann et al. 2004 for an opposite result). Based on this, the authors conclude that if some physical mechanism switches off the star formation when a galaxy enters a denser environment, it must do so on a short timescale, in order to preserve the shape of the EW(H$\alpha$) distribution.

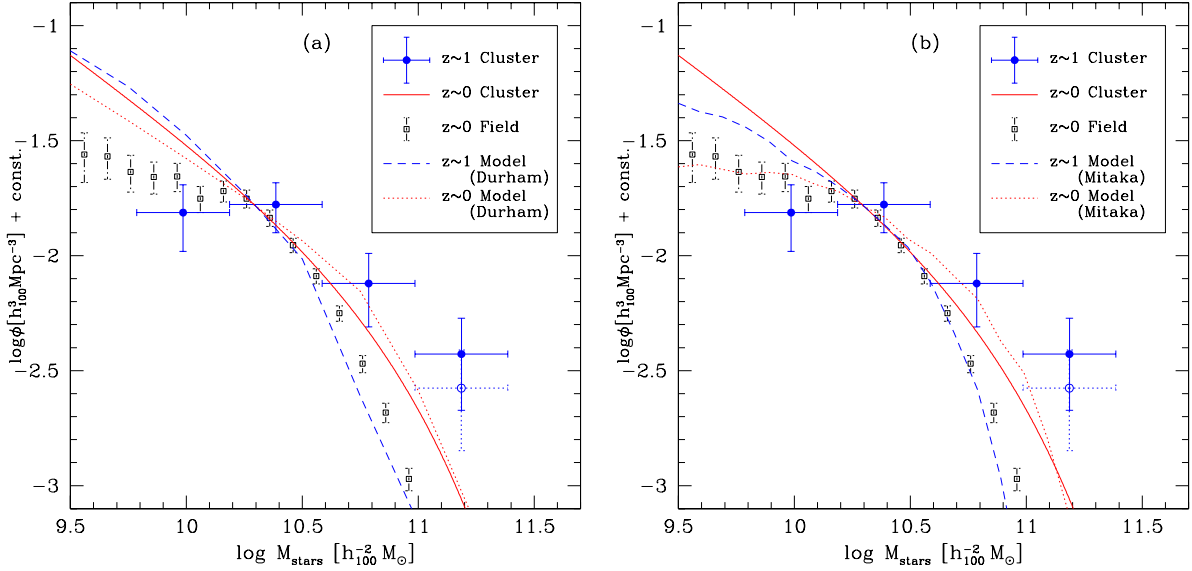

**Figure 6:** From Kodama & Bower (2003). The observed stellar mass functions at $z \sim 1$ and $z \sim 0$ compared to semi-analytic model predictions (Durham, left, and Mitaka, right).

The fact that a relation between star formation and density is observed also outside of clusters has often been interpreted as a sign that the environment starts affecting the star formation activity of galaxies (provoking a decline in star formation in galaxies that if isolated would continue forming stars) at relatively low densities, when a galaxy becomes part of a group, and strangulation has traditionally been invoked as the culprit. On the other hand, this appears in conflict with the fast timescale required for example by the H$\alpha$ and color studies mentioned above (Balogh et al. 2004a, 2004b).

Personally, I believe that the *existence* of such a correlation is more probably the result of a correlation between initial conditions (galaxy mass and/or local environment very early on, at the time the first stars formed in galaxies) and type of galaxy formed. The *exact shape* of the correlation, instead, is probably influenced by transformations happening in galaxies when they enter a different environment.

### 3.1 Gas content and gas/SF distribution within galaxies

In order to understand what happens to galaxies in clusters, two crucial pieces of information are 1) the gas content of cluster galaxies and 2) the spatial distribution of the gas and of the star formation activity within each galaxy.

It has been several years since it became evident that many spirals in clusters are deficient in HI gas compared to similar galaxies in the field (Giovanelli & Haynes 1985, Cayatte et al. 1990, see van Gorkom 2004 for a review). Most (but not all) of the HI deficient spirals are found at small distances from the cluster centre. In the central regions of clusters, the sizes of the HI disks are smaller than the optical disks, and a spatial displacement between the HI and the optical occurs in several cases (Bravo-Alfaro et al. 2000). The fraction of HI–deficient spirals increases going towards the cluster centre, and a correlation is observed between deficiency and orbital parameters: more deficient galaxies tend to be on radial orbits (Solanes et al. 2001).

All of these findings strongly suggest that ram pressure stripping, or at least gas stripping in general, plays an important role. On the other hand, the work from Solanes et al. (2001) has unexpectedly shown that the HI deficiency is observed out to 2 Abell radii. This result has raised the question whether the origin of the HI deficiency in the cluster outskirts can be consistent with the ram pressure scenario and whether can be simply due to effects such as large distance errors or rebounding at large clustercentric distances of galaxies that have gone through the cluster center (Balogh, Navarro & Morris 2000, Mamon et al. 2004, Moore et al. 2004, Sanchis et al. 2004).

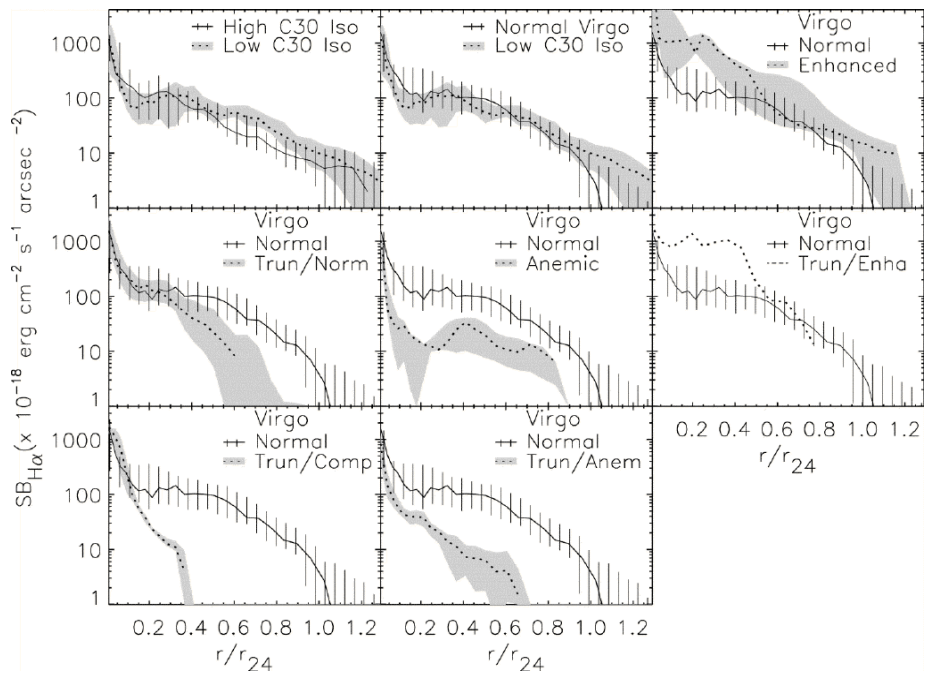

**Figure 7:** From Koopmann & Kenney 2004. Median Hα radial profiles for galaxies grouped in different classes according to their Hα distribution. The first two panels show profiles that can be considered "normal" (reproduced also in other panels). Truncated, anemic, enhanced classes are shown in all the other panels.

Recent works have yielded a census of the spatial distribution of the star formation (as observed in Hα emission) within cluster galaxies. These works have shown that the majority of the cluster galaxies have peculiar Hα morphologies, compared to field galaxies (Moss & Whittle 1993, 2000, Koopmann & Kenney 2004, Vogt et al. 2004). More than half of the spirals in the Virgo cluster, for example, have Hα radial profiles truncated from a certain radius on, while others have Hα suppressed throughout the disk, and in some cases, enhanced (Fig. 7) (Koopmann & Kenney 2004). On a sample of 18 nearby clusters, similar classes of objects are observed: spirals with truncated Hα emission and HI gas on the leading edge of the disk; spirals stripped of their HI with their star formation confined to the inner regions; and quenched spirals, in which the star formation is suppressed throughout the disk (Vogt et al. 2004). From these works, gas stripping appears a very important factor in determining both the gas content and the star formation activity of cluster spirals, though tidal effects are also found to be significant (Moss & Whittle 2000, Koopmann & Kenney 2004). Spirals in clusters thus appear to be heavily affected by the environment, and to be observed in different stages of their likely transformation from infalling star-forming spirals to cluster S0s.

It is apparently hard to reconcile the peculiar H$\alpha$ morphologies of spirals in nearby clusters with the fact that the distribution of H$\alpha$ equivalent widths (EWs) for blue galaxies in the 2dF and Sloan does not seem to vary significantly in clusters, groups and field (Fig. 9 in Balogh et al. 2004). This apparent contradiction still awaits an explanation.

## 4. Observations versus galaxy formation and evolution models

By now, mostly thanks to the scientific debate of the latest years, it is widespread wisdom that we are investigating a three-dimensional space, whose axis are redshift, environment and galaxy mass. In fact, the evolutionary histories and, in particular, the star formation activity of galaxies have similar (increasing) trends as a function of (higher) redshift, (lower) galaxy mass and (lower) density/mass of the environment. While observationally we are beginning to fill this 3-parameter space, the great theoretical challenge is to help comprehend from a physical point of view why this is the history of our Universe.

Given the topic of this School and meeting, I think it is appropriate to summarize what are the most important observational findings for simulations and models to address. Theory can be fundamental to:

1. explain the origin of the Morphology-Density relation; explain the cause of its evolution with redshift and the reason of its variation with environment;

2. reproduce the star formation history of galaxies as a function of the environment, in particular: the SF increase with lookback time, the existence of post-starburst galaxies in clusters, the origin and (predict the) evolution of the SF-Density relation

3. reproduce and explain the origin of the downsizing effect and the history of mass assembly

Hierarchical simulations show a correlation between color/morphology and density which is in qualitative agreement with observations (Kauffmann et al. 1999, Benson et al. 2001, and in prep., Diaferio et al. 2001, Springel et al. 2001) and with the Butcher-Oemler effect (Bower 1991, Kauffmann et al. 1995a, 1995b).

Let's take as example the morphology-density relation: a morphological segregation is a prediction of CDM simulations of large scale structure and semi-analytic models because the local density of galaxies and DM is related to the epoch of initial collapse: the most massive structures at any epoch are the earliest that collapsed. Even models that only include mergers and neglect any other environmental process obtain a trend in qualitative agreement with the MD relation: a morphological segregation is built-in at a very fundamental level in hierarchical theories of galaxy formation.

However, all models fail to reproduce the S0 population (which, it is worth reminding, represents 40% of the galaxy populations of rich clusters at low z), recognizing that additional processes seem to be required (Diaferio et al. 2001, Springel et al. 2001, Okamoto & Nagashima 2001, 2003, Benson et al. in prep.).

Another good example are the early-type galaxies. Observations point to a very heterogeneous evolution for those galaxies that today are passive red early-type galaxies in clusters, but *not* in the sense naively expected from models: the massive ones (for which mergers should be more important according to models) observationally appear truly old and homogeneous. It is by now established by an overwhelming number of observations that the more massive galaxies are, the older they are, thus that the age of galaxies have an anti-hierarchical behaviour with mass. At this meeting we have heard several lectures and talks reminding us of this observational evidence and of galaxy formation and evolution models reproducing this anti-hierarchical evolution with opportune assumptions (see e.g. Danese's and Cristiani's contributions to these proceedings, Chiosi & Carraro 2002). As we have seen in Sec. 2.2 the mass assembly of galaxies in clusters is another field where until now there is a mismatch between models and observations. But the comparison between galaxy models within a CDM scenario and observations of galaxies in clusters is still largely unprobed territory, and a fruitful interplay can be expected in the coming years, as promised by the progress shown at this meeting.